\documentclass{article}
\usepackage{spconf,amsmath,graphicx}
\usepackage{booktabs} 
\usepackage{multirow}
\usepackage{hyperref}
\usepackage{threeparttable}
\title{The RoyalFlush Automatic Speech Diarization and Recognition System for In-Car Multi-Channel Automatic Speech Recognition Challenge}

\name{Jingguang Tian$^{*}$, Shuaishuai Ye$^{*}$, Shunfei Chen, Yang Xiang, Zhaohui Yin, Xinhui Hu, Xinkang Xu}
\address{Hithink RoyalFlush AI Research Institute, Zhejiang, China}

\begin{document}
\ninept
\maketitle
\def\thefootnote{*}\footnotetext{Equal contribution.}\def\thefootnote{\arabic{footnote}}
\begin{abstract}
\vspace{-0.15cm}
This paper presents our system submission for the In-Car Multi-Channel Automatic Speech Recognition (ICMC-ASR) Challenge, which focuses on speaker diarization and speech recognition in complex multi-speaker scenarios. To address these challenges, we develop end-to-end speaker diarization models that notably decrease the diarization error rate (DER) by 49.58\% compared to the official baseline on the development set. For speech recognition, we utilize self-supervised learning representations to train end-to-end ASR models. By integrating these models, we achieve a character error rate (CER) of 16.93\% on the track 1 evaluation set, and a concatenated minimum permutation character error rate (cpCER) of 25.88\% on the track 2 evaluation set.
\end{abstract}
\vspace{-0.1cm}

\vspace{-0.2cm}
\section{Introduction}
\label{sec:intro}

As cars have become an essential part of daily life, speech-based interaction in cars, a natural method of human-computer interaction, is gaining prominence. Unlike ASR systems used at home or in meetings, in-car systems face unique challenges. These challenges stem from the complex acoustic environment of the cockpit, various noises from both inside and outside the car, and different driving conditions. This has led to the establishment of the ICMC-ASR competition \footnote{https://icmcasr.org/}. The challenge is split into two tracks: track 1 focuses on single-speaker ASR with oracle segmentation, while track 2 concentrates on multi-speaker ASR, requiring participants to engineer systems for speaker diarization and transcription.

\vspace{-0.2cm}
\section{System Description}
\label{sec:system description}

\begin{figure}[t]
\vspace{-0.5cm}
\begin{minipage}[b]{.4\linewidth}
  \centering
  \centerline{\includegraphics[width=3.5cm]{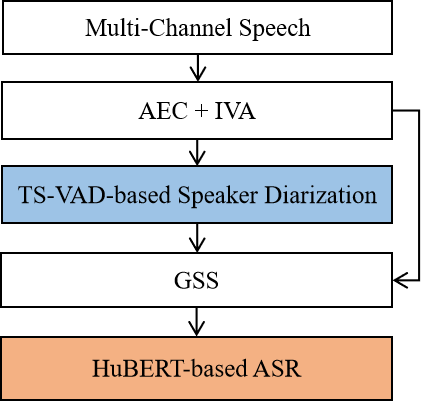}}
  \vspace{-0.1cm}
  \centerline{(a)}\medskip
\end{minipage}
\begin{minipage}[b]{.6\linewidth}
  \centering
  \centerline{\includegraphics[width=4.6cm]{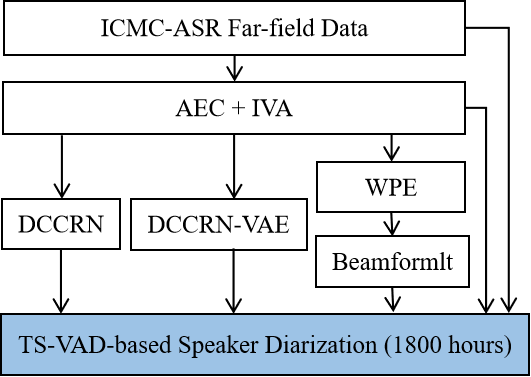}}
  \vspace{-0.1cm}
  \centerline{(b)}\medskip
\end{minipage}
\hfill
\begin{minipage}[b]{1.0\linewidth}
  \centering
  \centerline{\includegraphics[width=6cm]{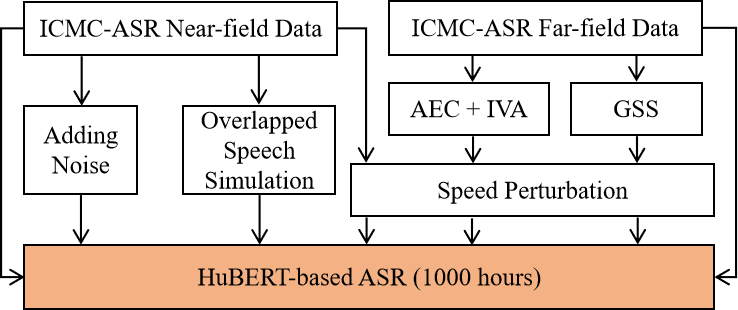}}
  \vspace{-0.1cm}
  \centerline{(c)}\medskip
\end{minipage}
\vspace{-0.82cm}
\caption{(a) The overview of the ASDR system; (b) Data flow for speaker diarization training; (c) Data flow for ASR training.}
\label{fig:the overview of system}
\vspace{-0.7cm}
\end{figure}

Fig. \ref{fig:the overview of system} (a) illustrates our modular automatic speech diarization and recognition (ASDR) system for the ICMC-ASR challenge. Multi-channel speech signals undergo processing by acoustic echo cancellation (AEC) and independent vector analysis (IVA) to yield enhanced audio. The speaker diarization module primarily relies on target-speaker voice activity detection (TS-VAD). Guided source separation (GSS) \cite{raj2022gpu} utilizes speaker activity information provided by TS-VAD and enhanced audio to perform front-end enhancement of overlapped speech signals. Following GSS, we partition the audio based on the diarization output and feed it into a single-speaker HuBERT-based ASR module.

\vspace{-0.2cm}
\subsection{Speaker Diarization}
TS-VAD was introduced as a method to identify the vocal activities of a specific group of speakers within an input audio signal, using the known profiles of these speakers \cite{medennikov20}. Typically, these speaker profiles are acquired in advance through a process known as clustering-based speaker diarization. Our model \cite{tian2022royalflush} for this process consists of TCN-based VAD \cite{yin2023}, a Resnet34-TSDP speaker embedding extractor, and NMESC. In contrast to the original TS-VAD, which uses i-vector for target-speaker embedding, we employ neural speaker embedding extracted by Resnet34-TSDP for target-speaker detection. The TS-VAD model is composed of three modules: 1) a frame-level speaker embedding encoder, which has the same architecture as the target-speaker embedding extractor, 2) a speaker detection module based on either Transformer or Conformer, 3) a combining module that utilizes either BLSTM or DPRNN \cite{luo2020dual}.
\vspace{-0.2cm}
\subsection{Speech Recognition}
Drawing inspiration from IRIS \cite{chang2022end}, we developed an end-to-end speech recognition system. This system comprises three key components: a DCCRN-VAE \cite{xiang2024deep} for speech enhancement (SE), a HuBERT \cite{2021HuBERT} for self-supervised learning representation (SSLR), and a joint CTC/attention-based encoder-decoder for ASR. The acoustic model of the ASR mirrors the official baseline, utilizing an E-Branchformer-based encoder and a transformer-based decoder. Additionally, we incorporated a transformer-based language model (LM) into our system.
\begin{table*}[th]
\vspace{-0.85cm}
\setlength\tabcolsep{0.15cm}
\caption{The results of the ASDR system on the ICMC-ASR development and evaluation set.}
\label{table1}
\centering
\begin{tabular}{ccccccc}
\hline
\multicolumn{4}{c}{Development Set}                                  & \multicolumn{3}{c}{Evaluation set} \\
\cmidrule(r){1-4} \cmidrule(r){5-7}
Speaker Diarization Model      & DER    & ASR Model        & CER     & ASDR Model & Track 1 (CER)   & Track 2 (cpCER)  \\ \hline
Baseline          & 54.79\% & Baseline            & 32.92\% & Baseline & 26.24\%  & 72.88\%          \\
Transformer-BLSTM & 5.90\%  & HuBERT-ASR                    & 23.56\% & -         & -    & -                \\
Transformer-DPRNN & 5.67\%  & SE-HuBERT-ASR-Part & 23.07\% & -          & -    & -                \\
Conformer-BLSTM   & 5.95\%  & SE-HuBERT-ASR-All   & 22.85\% & -          & -    & -                \\
Conformer-DPRNN   & 5.69\%  & SE-HuBERT-ASR-All-LM   & 22.84\% & -       & -    & -                \\
Fusion            & \textbf{5.21}\%  & Fusion        & \textbf{21.77}\%  & Fusion  & \textbf{16.93}\%  & \textbf{25.88}\%  \\ \hline
\vspace{-0.85cm}
\end{tabular}
\end{table*}

\vspace{-0.2cm}
\section{Experimental Setup and Results}
\label{sec:Exp setup}

\vspace{-0.2cm}
\subsection{Datasets}
Besides the ICMC-ASR corpus, we incorporated external data as well. A comprehensive presentation of the training data can be found in Table \ref{table2}. The data flow of each training process is depicted in Fig. \ref{fig:the overview of system} (b) and (c). We generated a substantial amount of training data using a diverse range of data processing methods. These include DCCRN and DCCRN-VAE-based SE, WPE \cite{tian2022royalflush}, BeamformIt \cite{tian2022royalflush}, adding noise, overlapped speech simulation, GSS-based augmentation, and speed perturbation.
\begin{table}[ht]
\vspace{-0.5cm}
\setlength\tabcolsep{0.03cm}
\caption{The data utilized for training in our system.}
\label{table2}
\centering
\begin{tabular}{cc}
\hline
Model         & Training data           \\ \hline
TCN \& TS-VAD  & ICMC-ASR far-field training set \\
Resnet34-TSDP & CNCeleb \cite{li2022cn}      \\
DCCRN-VAE \& DCCRN   & DNS \cite{xiang2024deep}  \\
HuBERT        & WenetSpeech \cite{zhang2022wenetspeech}  \\
HuBERT-ASR    & ICMC-ASR near and far-field training set \\ \hline
\end{tabular}
\vspace{-0.5cm}
\end{table}

\vspace{-0.2cm}
\subsection{System Setup}
\vspace{-0.2cm}
\subsubsection{Speaker Diarization}
For speaker diarization, we created four TS-VAD models, each uniquely integrating speaker detection and combining modules. The models were Transformer-BLSTM, Transformer-DPRNN, Conformer-BLSTM, and Conformer-DPRNN. Data augmentation, which involves adding noise, is performed on-the-fly using ICMC-ASR noise audio. In the inference stage, the TS-VAD results from different single-channel speech are fused using DOVER-Lap \cite{raj2021dover}. Several other aspects of the model bear similarity to \cite{wang2022cross}.

\vspace{-0.2cm}
\subsubsection{Speech Recognition}
We adopted a pre-training and fine-tuning scheme to train ASR models \cite{chang2022end}. We utilized a total of four ASR models, namely HuBERT-ASR (without SE), SE-HuBERT-ASR-Part (fine-tuning only SE and ASR), SE-HuBERT-ASR-All (jointly fine-tuning all parameters), and SE-HuBERT-ASR-All-LM (incorporating a LM). During the inference stage, the results from different ASR models are fused using ROVER \cite{rover}.
\vspace{-0.2cm}
\subsection{Experimental Results}
\label{sec:results}
\vspace{-0.2cm}
\subsubsection{Speaker Diarization}
The performances of the speaker diarization models on the development set are shown in Table \ref{table1}. The four TS-VAD models exhibit comparable performance. However, when used as a combining module, DPRNN holds a marginal edge over BLSTM. As for speaker detection modules, Transformer and Conformer show negligible differences. Fusing the four TS-VAD models leads to a 5.21\% DER, which is a 49.58\% absolute reduction over the official baseline.

\vspace{-0.2cm}
\subsubsection{Speech Recognition}
As shown in Table \ref{table1}, the SE-HuBERT-ASR-All-LM achieves the best performance among single models. Compared with the official baseline, the fusion scheme results in an absolute CER reduction of 11.15\% (from 32.92\% to 21.77\%) on the development set and 9.31\% (from 26.24\% to 16.93\%) on the evaluation set for track1. By integrating the results from both the fused TS-VAD models and the fused ASR models, we obtained an absolute cpCER reduction of 47.00\% (from 72.88\% to 25.88\%) on the evaluation set of track2.

\vspace{-0.2cm}
\section{Conclusions}
\label{sec:conclusions}

We introduced the ASDR system for the ICMC-ASR challenge. Our findings highlight the significant contribution of TS-VAD-based speaker diarization to this task. Furthermore, we demonstrate that a robust speech recognition system can be constructed under noisy conditions by collectively fine-tuning SE, SSLR, and ASR.

% Below is an example of how to insert images. Delete the ``\vspace'' line,
% uncomment the preceding line ``\centerline...'' and replace ``imageX.ps''
% with a suitable PostScript file name.
% -------------------------------------------------------------------------

% To start a new column (but not a new page) and help balance the last-page
% column length use \vfill\pagebreak.
% -------------------------------------------------------------------------
%\vfill
%\pagebreak

%\vfill\pagebreak

% References should be produced using the bibtex program from suitable
% BiBTeX files (here: strings, refs, manuals). The IEEEbib.bst bibliography
% style file from IEEE produces unsorted bibliography list.
% -------------------------------------------------------------------------

\vspace{-0.2cm}
%\scriptsize
%\ninept
\bibliographystyle{IEEEbib}
\bibliography{strings,refs}

\begin{thebibliography}{10}

\bibitem{raj2022gpu}
D~Raj, D~Povey, and S~Khudanpur,
\newblock ``Gpu-accelerated guided source separation for meeting transcription,''
\newblock {\em arXiv preprint arXiv:2212.05271}, 2022.

\bibitem{medennikov20}
I~Medennikov, M~Korenevsky, and et~al.,
\newblock ``Target-speaker voice activity detection: A novel approach for multi-speaker diarization in a dinner party scenario,''
\newblock in {\em Interspeech}, 2020, pp. 274--278.

\bibitem{tian2022royalflush}
J~Tian, X~Hu, and X~Xu,
\newblock ``Royalflush speaker diarization system for icassp 2022 multi-channel multi-party meeting transcription challenge,''
\newblock {\em arXiv preprint arXiv:2202.04814}, 2022.

\bibitem{yin2023}
Z~Yin, J~Tian, and et~al.,
\newblock ``Large-scale learning on overlapped speech detection: New benchmark and new general system,''
\newblock {\em arXiv preprint arXiv:2308.05987}, 2023.

\bibitem{luo2020dual}
Y~Luo, Z~Chen, and T~Yoshioka,
\newblock ``Dual-path rnn: efficient long sequence modeling for time-domain single-channel speech separation,''
\newblock in {\em ICASSP}. IEEE, 2020, pp. 46--50.

\bibitem{chang2022end}
X~Chang, T~Maekaku, and et~al.,
\newblock ``End-to-end integration of speech recognition, speech enhancement, and self-supervised learning representation,''
\newblock in {\em INTERSPEECH}, 2022, vol. 2022, pp. 3819--3823.

\bibitem{xiang2024deep}
Y~Xiang, J~Tian, and et~al.,
\newblock ``A deep representation learning-based speech enhancement method using complex convolution recurrent variational autoencoder,''
\newblock in {\em ICASSP}. IEEE, 2024.

\bibitem{2021HuBERT}
W.-N Hsu, Y.-H.~H Tsai, and et~al.,
\newblock ``Hubert: How much can a bad teacher benefit asr pre-training?,''
\newblock in {\em ICASSP}, 2021, pp. 6533--6537.

\bibitem{li2022cn}
L~Li, R~Liu, and et~al.,
\newblock ``Cn-celeb: multi-genre speaker recognition,''
\newblock {\em Speech Communication}, vol. 137, pp. 77--91, 2022.

\bibitem{zhang2022wenetspeech}
B~Zhang, H~Lv, and et~al.,
\newblock ``Wenetspeech: A 10000+ hours multi-domain mandarin corpus for speech recognition,''
\newblock in {\em ICASSP}. IEEE, 2022, pp. 6182--6186.

\bibitem{raj2021dover}
D~Raj, L.~P Garcia-Perera, and et~al.,
\newblock ``Dover-lap: A method for combining overlap-aware diarization outputs,''
\newblock in {\em SLT}. IEEE, 2021, pp. 881--888.

\bibitem{wang2022cross}
W~Wang, X~Qin, and M~Li,
\newblock ``Cross-channel attention-based target speaker voice activity detection: Experimental results for the m2met challenge,''
\newblock in {\em ICASSP}. IEEE, 2022, pp. 9171--9175.

\bibitem{rover}
J.~G Fiscus,
\newblock ``A post-processing system to yield reduced word error rates: Recognizer output voting error reduction (rover),''
\newblock in {\em ASRU}, 1997.

\end{thebibliography}

\end{document}